# Ultra-fast Double Pulse All-Optical Re-switching of a Ferrimagnet

C. Banerjee, K. Rode, G. Atcheson, S. Lenne, P. Stamenov, J. M. D. Coey and J. Besbas*

CRANN, AMBER and School of Physics, Trinity College Dublin, Dublin 2, Ireland

*besbasj@tcd.ie

**Abstract**
All-optical re-switching has been investigated in the half-metallic Heusler ferrimagnet $Mn_2Ru_{0.9}Ga$, where Mn atoms occupy two inequivalent sites in the XA-type structure. The effect of a second 200 fs 800 nm pump pulse that follows a first pulse, when both are above the threshold for switching, is studied as a function of $t_{12}$, the time between them. The aims are to identify the physical mechanisms involved and to determine the minimum time needed for re-switching. The time trajectory of the switching process on a plot of sublattice angular momentum, $S^{4a}$ vs $S^{4c}$, is in three stages; When t < 0.1 ps, the sublattice moments are rapidly disordered, but not destroyed, while conserving net angular momentum via optical spin-wave excitations. This leads to transient parallel alignment of the residual Mn spins in the first quadrant. The net angular momentum associated with the majority sublattice then flips in about 2 ps, and a fully-reversed ferrimagnetic state is then established via the spin-lattice interaction, which allows re-switching provided $t_{12}$ > 10 ps.

Single-pulse all-optical switching of magnetization (SP-AOS) is of both fundamental and technological interest [1–3]. Despite intense scrutiny over the last two decades, the microscopic origin of the effect is still poorly understood, but the possibility of switching the magnetisation of a thin film between two stable states on a picosecond timescale without recourse to an external magnetic field is intriguing and technologically relevant in the quest for ever-faster and more energy-efficient information technologies [4–6]. Here we establish the minimum time that must elapse between two pulses, if the second one is to re-establish the original state. Our results advance the fundamental understanding of SP-AOS and highlight its potential for future application in technology.

We have recently shown that the near-cubic XA-ordered (F-43m) ferrimagnetic Heusler alloy $Mn_2Ru_xGa$ (MRG) exhibits SP-AOS [7], and that switching is driven by antiferromagnetic exchange between the crystallographically-inequivalent $4a$ and $4c$ Mn sublattices [8]. The inequivalence is the source of two key properties of MRG. First, the states close to the Fermi level are associated predominantly with one of the sublattices, which we identified in compounds with $x \approx 0.7$ as $4c$ [9], resulting in half-metallic character. This sublattice dominates magneto-optic Kerr effect (MOKE) [10]. All MOKE-based measurements shall therefore be understood as reflecting the response of the $4c$ sublattice. Second, due to the hierarchy of the intra- and inter-sublattice exchange constants, $J_{aa} > -J_{ac} > |J_{cc}|$ [11], the $4c$ sublattice exhibits the higher moment at $T = 0$ K, but its magnitude falls faster with temperature than that of $4a$ so that magnetic compensation occurs at a temperature $T_{comp}$ where the two sublattice magnetizations are equal but opposite in sign [9,12]. We found that SP-AOS is only possible below $T_{comp}$, when at equilibrium the absolute value of the $z$-projection of the angular momentum of $4c$ manganese exceeds that of $4a$ manganese [7,8].

The MRG sample studied here, $Mn_2Ru_{0.9}Ga$, was grown by DC magnetron co-sputtering from $Mn_2Ga$ and Ru targets on MgO (001) single-crystal substrates heated to 425°C using an ultra-high vacuum DCA multi-chamber deposition and characterisation tool (Trifolium Dubium, National Access Facility). The film was capped by 2 nm protective layer of naturally oxidized $AlO_x$, deposited at room temperature, followed by 8 nm of $SiO_2$. Biaxial, substrate-induced strain induces a slight tetragonal distortion of the cubic Heusler structure, resulting in perpendicular magnetocrystalline anisotropy of the film [13] and a room-temperature coercivity of 450 mT. The compensation temperature was found to be 469 K from a thermal scan of the remnant magnetization, measured by SQUID magnetometry on another sample prepared in identical conditions. Further details on the structural, magnetic, magneto-optic and magneto-transport properties of MRG can be found elsewhere [9,12,13].

200 fs laser pulses ($\lambda = 800$ nm) were sourced from a mode-locked Ti:sapphire-based laser system. The system was operated in single-pulse mode for *ex situ* imaging, whereas for stroboscopic time-resolved magnetisation dynamics, the pulse repetition rate was 1 kHz. A portion of the beam was used for second harmonic generation in a β-barium borate crystal creating the probe beam ($\lambda = 400$ nm). Its delay with respect to a pump beam, was adjusted by a mechanical translation stage. A pair of pulses with variable delay were generated from a single pulse using a Michelson interferometer on the pump beam path with one arm mounted on a mechanical translation stage. MOKE imagery was recorded *ex situ* after exposure using an EVICO Kerr microscope with red light in zero applied magnetic field. For all stroboscopic measurements, an applied field of 950 mT was applied perpendicular to the sample surface using an electromagnet.

Figure 1(a) illustrates the 'toggle' nature of SP-AOS. After the first pump exposure, the irradiated spot reverses its magnetisation, and subsequent pulses toggle the magnetisation back and forth. We also show MOKE micrographs for varying pump powers (Fig. 1(b)) from which the Gaussian pump beam diameter and the threshold for switching were determined using the Liu method [14]. We find a threshold of 3.5 mJ cm$^{-2}$ and a spot size of about 190 µm.

In Figure 1(c) we plot the time evolution of the MOKE after a single pump of 8.9 mJ cm$^{-2}$, well above threshold. The solid line is a bi-exponential fit to the data with characteristic times 100 fs and 1.9 ps. Since our probe pulse has duration ~200 fs, while the pump is slightly stretched to ~250 fs, due to additional optical elements in the beam path, our time resolution close to the pump is ~325 fs. The two characteristic times are in agreement with our understanding of the SP-AOS process in MRG: Immediately after the pump, the two sublattices demagnetise while conserving net angular momentum such that $dS_z^{4a}/dt = - dS_z^{4c}/dt$ [7,8]. This step is governed by the inter-sublattice exchange, and it leads to a state where the average z-projections of the two sublattice moments are aligned parallel because $|S_z^{4c}| > |S_z^{4a}|$. This is referred to as the transient 'ferromagnetic-like' state [1], and it is a necessary but not sufficient condition for switching [15]. The associated time scale is ~150 fs for Gd(FeCo)$_3$ [1] and ~50 fs for MRG on account of the ~3 times stronger intersublattice exchange constant in the manganese alloy [11]. We infer that for times $t \leq 325$ fs, the 4a sublattice has switched its orientation while 4c has not. At longer times, $t > 325$ fs, a second process becomes dominant. Angular momentum is no longer conserved, which allows the 4c sublattice to switch at $t \sim 1$ ps and a quasi-static state is reached at $t \sim 10$ ps, consistent with the spin-lattice relaxation time in MRG [8]. On a longer timescale of ~ 300 ps, the lattice cools down to near-ambient temperature by heat flow into the substrate.

Figure 2(a) illustrates the switching with two pump pulses. A first pulse at $t = 0$ reverses the magnetisation; a second pulse at $t_{12} = 110$ ps toggles it back. To confirm that magnetic switching actually occurred, we first recorded a MOKE field loop before any excitation (Fig. 2(b)), then one at $t = 15$ ps after a first pump pulse (Fig. 2(c)), and another at $t = 285$ ps, after both (Fig. 2(d)). The sign reversal of the loops confirms the magnetic switching.

We then determine the minimum value of $t_{12}$ that allows the second pulse to toggle the magnetisation. Figure 3(a) shows MOKE micrographs after the sample has been irradiated with two pulses of 4.1 mJ cm$^{-2}$, separated by $t_{12}$ = 9, 11, 11.7 or 12 ps. The first pulse switches the area where the intensity of the Gaussian beam exceeds the threshold at room temperature, and also increases the lattice temperature by approximately 65 K in about 2 ps [7]. This increased temperature decays slowly by heat flow to the substrate. As the threshold fluence decreases with increasing temperature (decreasing net magnetisation), the second pulse toggles an area that is *bigger* than the first. This is clearly visible in Fig. 3(a) for 12 ps pump separation: the central bright spot was switched once by the first pump, then toggled back again by the second, while the dark ring surrounding it is unchanged magnetically by the first and switched by the second. The threshold fluence at this transient higher temperature (365 K) is only 2.9 mJ cm$^{-2}$, determined from the ratio of the toggled areas.

Re-switching does not occur at a pump separation of 9 ps, whereas at $t_{12}$ = 12 ps it is complete. For pump separations of 11.0 and 11.7 ps we find a third central region where re-switching was not achieved because the higher peak pump intensity requires a longer time to reach equilibrium, even though the relevant time constants are the same. This is illustrated in Fig. 3(a) where we increase the fluence of the first pump to 8.2 mJ cm$^{-2}$ and the second to 6.1 mJ cm$^{-2}$. For these fluences, the areas switched by the first pump and re-switched by the second are nearly equal, and the central non-reswitching area remains visible up to a pump separation of 70 ps. The results are summarized in Fig. 3(c) where we show the re-switched fraction as a function of pump separation $t_{12}$ for fluences of 4.1, 4.8 and 8.2 mJ cm$^{-2}$. We highlight two points in the data. First, the lattice temperature does *not* need to exceed $T_{comp}$ to ensure switching, as is observed in Gd(FeCo)$_3$ − excessive heating actually prevents re-switching. Second, the fundamental limit on repetition rate is not uniquely determined by the heat transfer to the substrate. The relevant time is the spin-lattice relaxation time, the time needed for magnetic damping.

Based on the original studies of amorphous Gd(FeCo)$_3$ [1,2], SP-AOS was believed to depend on two conditions. First, it was thought that the demagnetisation times of the two sublattices needed to be substantially different, so that the *z*-projections of their moments could cross zero at different times. Second, it was thought that high spin polarisation inhibited efficient demagnetisation. These expectations were overturned by our observation of switching in MRG. There, two sublattices composed of the same element would be expected to demagnetize at similar rates. Furthermore, although overwhelmingly one of the sublattices contributes the majority of the states at the Fermi level, the material nevertheless exhibits SP-AOS. We now discuss the situation in light of our new findings.

The on-atom Coulomb interaction integrals for $3d^5$ manganese (Slater $F^2$ and $F^4$) are 0.6 and 0.4 Ry (8.2 and 5.4 eV) and the first thermally excited configuration is $(5/7 - 25/49)$ $F^2 + (5/7 - 190/441)$ $F^4$ higher in energy, corresponding to an energy of 3.2 eV [16,17]. For MRG we infer that the atomic moment and the exchange integrals remain, to a very good approximation, time independent. The corresponding energies are 3.5, 0.35, 0.14 and 0.07 eV for Gd ($4f^7$), Tb ($4f^9$), Co ($3d^5$) and Fe ($3d^6$) respectively, suggesting that this will not be the case for Tb, Co and Fe [18]. Optically-induced transitions to excited states do not change S due to the magneto-optical selection rules [16]. We must therefore discuss our findings in the language of spin waves and precession [19], noting however that the usual models for spin waves assume that the *x*- and *y*-projections of the atomic moments are small ($S_z \gg S_x, S_y$), an assumption that is clearly invalid for SP-AOS.

Ferrimagnets exhibit at least two orthogonal spin wave modes if axial symmetry is unbroken. In one mode, the two sublattices precess together without changing the angle between them; in the other, they precess in antiphase. The two are frequently referred to as the 'acoustic' or 'ferromagnetic' and 'optical' or 'antiferromagnetic' modes, respectively. In amorphous Gd(FeCo)$_3$, axial symmetry is broken by structural inhomogeneity [19] (Gd tends to cluster) whereas in MRG non-collinearity of the ferrimagnetic ground state [20], four-fold sublattice-specific magnetocrystalline anisotropy of opposite signs [21], and preferential absorption by light of one sublattice play the same role. The intense electronic excitation provided by the pump pulse excites a multitude of magnons. In the absence of axial symmetry, the optical mode can be efficiently excited [22], leading to the first-quadrant 'ferromagnetic' aligned state discussed earlier. The relevant times are those associated with exchange energies via the uncertainty principle ~ 100 fs, which are comparable to the 200 fs duration of the pulse in our experiments [23]. We note that this process is fast because it conserves angular momentum. It only depends on the magnetic system absorbing the energy deposited by the pump. This is often called exchange scattering [24]. Thermodynamically, the maximum energy that can be absorbed by the magnetic system while conserving angular momentum leads to $S^{4c}_{z\,min} = (n_{aa} + n_{ac})/(n_{aa} + n_{cc} + 2n_{ac}) (S^{4a}_{z0} + S^{4c}_{z0})$, where $n$ ($n_{ij} > 0$) are the Weiss molecular field constants. $S_z^{4c}$ remains positive, while $S_z^{4a}$ has switched and the associated time is that needed for the first stage of switching.

Following this, the magnetic system loses energy by coupling to the lattice and the 4*c* sublattice reverses its magnetic polarity while the 4a, that has already changed polarity, is increasing. When the 4*c* crosses zero, the inter-sublattice exchange will align it antiparallel to 4*a*. This process does not conserve net angular momentum and probably requires emission of optical phonons. The experimentally determined timescale is 1.9 ps which represents the time needed for the second stage, as both sublattices are now antiparallel to their initial directions. We believe this second stage is driven by continued demagnetisation of the 4$c$ sublattice. It is telling that the threshold for switching decreases when the temperature increases towards $T_{comp}$, as the residual z-projection at $t$ = 325 fs is reduced. We speculate that very close to $T_{comp}$ the threshold fluence for SP-AOS could be very substantially reduced, albeit only for extremely short pump pulses [8].

Lastly, the optical magnons scatter into the long wavelength acoustic modes of frequency ~ 100 GHz [23] and are damped on the spin-lattice relaxation timescale (~10 ps) when $S_z^{4c}$ regains a higher magnitude than $S_z^{4a}$, unless the lattice has already heated above $T_{comp}$. This is the time that finally marks the completion of the magnetic reversal, and it then becomes possible to repeat the process and toggle the magnetisation with a second pulse. The half metallicity of MRG is beneficial, as it will increase the damping of the 4$c$ sublattice due to Fermi surface breathing [25] and allow it to relax faster than 4$a$, decreasing the time that must elapse between subsequent toggle events. The whole three-stage process is illustrated on Fig. 4 by the track from initial to final states, where the relevant times after the pump are marked on a logarithmic scale on the red trajectory. Plausible spin configurations at different times are illustrated in the inset. The four-quadrant representation of a two-sublattice magnet in Fig. 4 has been used by Mentink *et al.* [26], originally inspired by Bar'yakhtar [24].

Finally, we comment on the energy requirements for a potential application. We have shown that threshold fluences, as low as 2.9 mJ cm$^{-2}$ or 0.3 fJ, suffice to switch a (10 nm)³ element, assuming that 35% of the light is absorbed by a 30 nm thick MRG thin film. This is an order of magnitude more than current records for transparent magnetic insulators [3]. However, the metallic nature of MRG permits integration with other spin electronic circuitry, thereby creating an opportunity to bring the speed of optics to magnetism and electronics. Possible applications include beam steering using diffraction elements, such as Fresnel zone plates, where MRG (or some future material) forms the 'dark' elements. The focal point of the zone plate could thus be changed every 10 ps.

In conclusion, the relevant timescales for SP-AOS are the exchange time and the spin-lattice relaxation time, which we evaluate from our two-pulse experiments. We infer that SP-AOS requires axial symmetry breaking, either by structural inhomogeneity, or by competition between magnetocrystalline anisotropy and exchange. The hierarchy of exchange constants in a ferrimagnet is critical to promote low-energy SP-AOS. Repeated toggle switching is envisaged at rates as high as 100 GHz, provided the lattice temperature remains below $T_{comp}$.

Reducing the spin-lattice relaxation time could increase this frequency. To our knowledge this is the fastest switch from one stable magnetic state to another ever observed.

**Acknowledgements.**

This work was supported by Science Foundation Ireland under contract 16/IA/4534 ZEMS and the European Union Horizon 2020 research and innovation grant agreement 737038 'TRANSPIRE'. Dr. C. Banerjee is grateful to the Irish Research Council for her postdoctoral fellowship. The work was carried out in the CRANN Photonics Laboratory, where we are grateful to Dr. Jing Jing Wang for technical support. The Trifolium Dubium deposition/characterisation platform was funded by Science Foundation Ireland under grant 15/RI/3218.

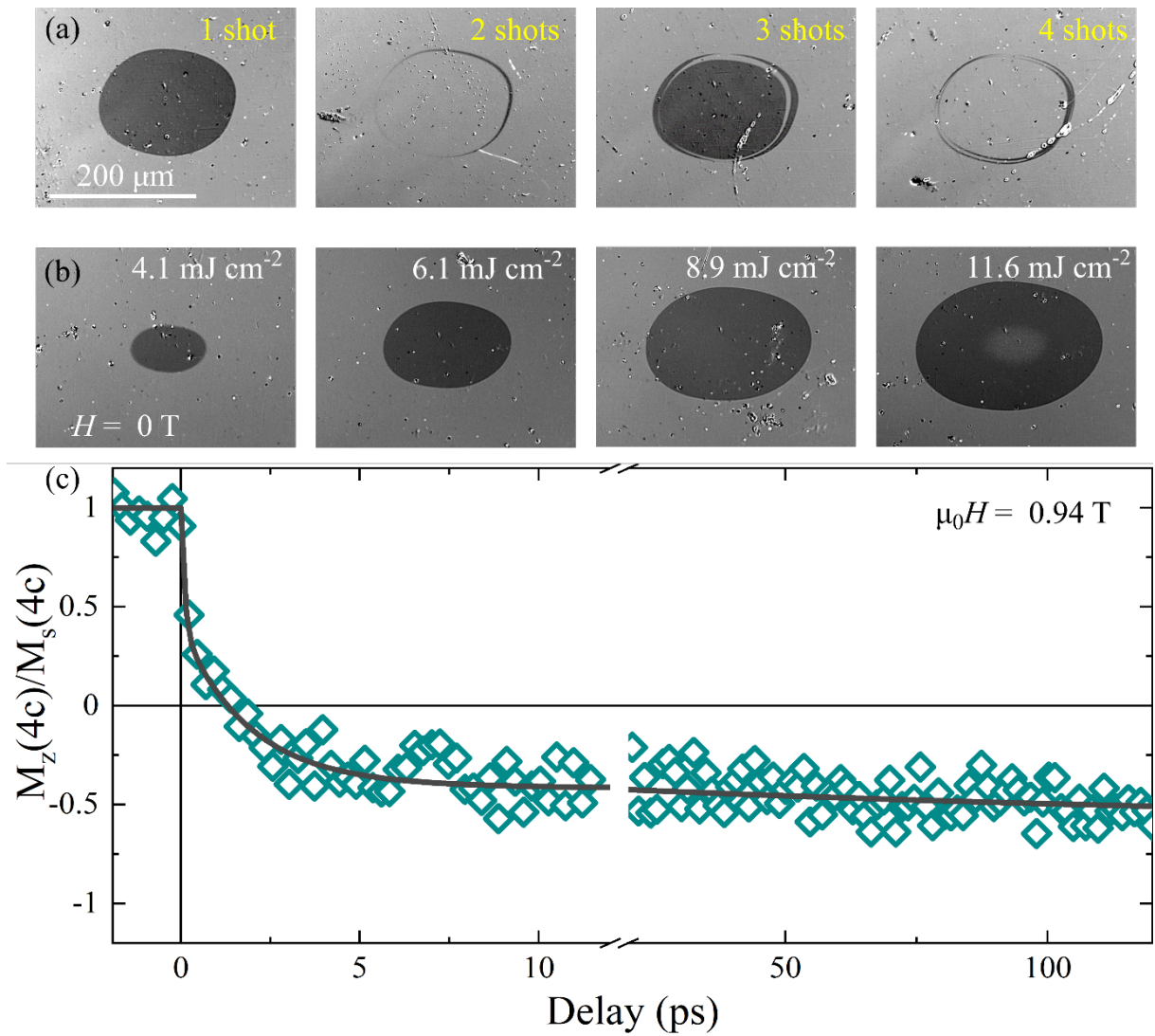

Figure 1: (a) All-optical toggle switching of magnetization in $Mn_2Ru_{0.9}Ga$ at a fluence of 8.2 mJ cm$^{-2}$. (b) Domain size as a function power. (c) Magnetization dynamics for a pump of fluence of 8.9 mJ cm$^{-2}$. The solid line is a guide to the eye based on three exponentials with characteristic times 100 fs, 1.9 ps and 320 ps.

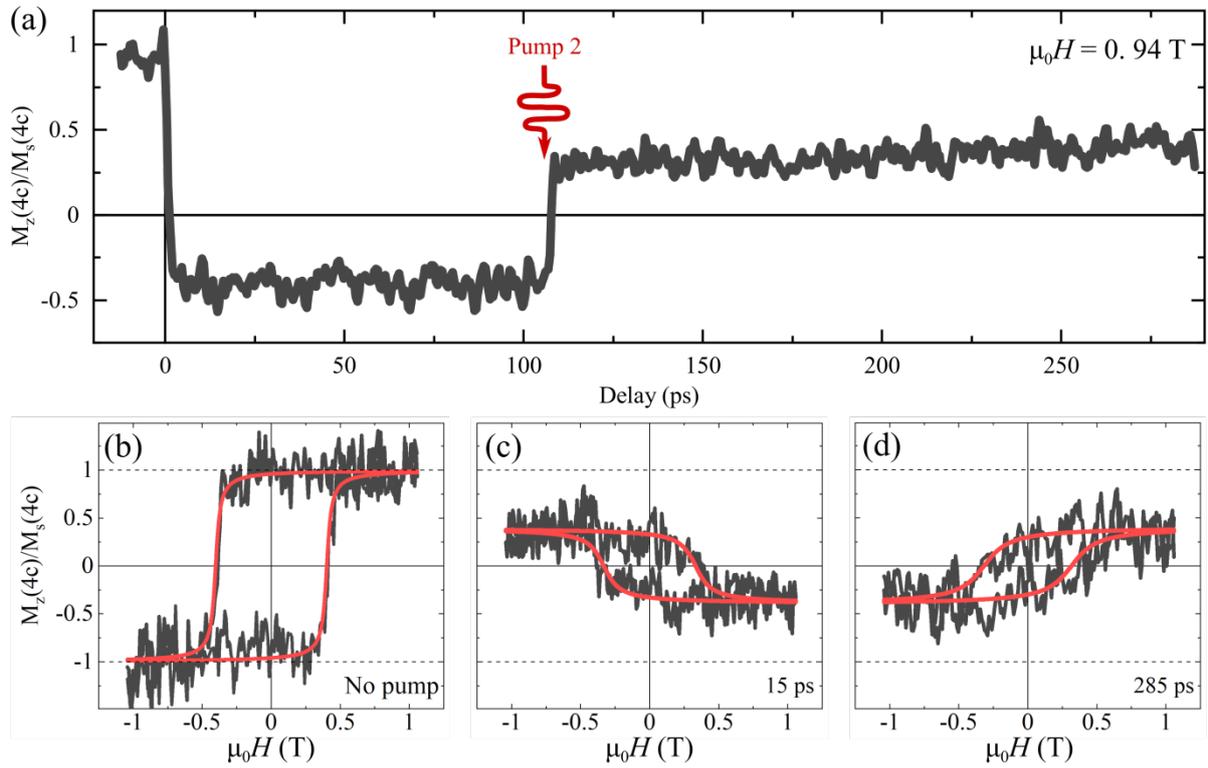

Figure 2: (a) Transient Kerr signal in presence of two pump pulses separated by 110 ps. (b) Hysteresis loop measured by the probe beam in absence of any pump excitation. (c) and (d) are field loops at delays $t_{12}$ = 15 ps and 285 ps.

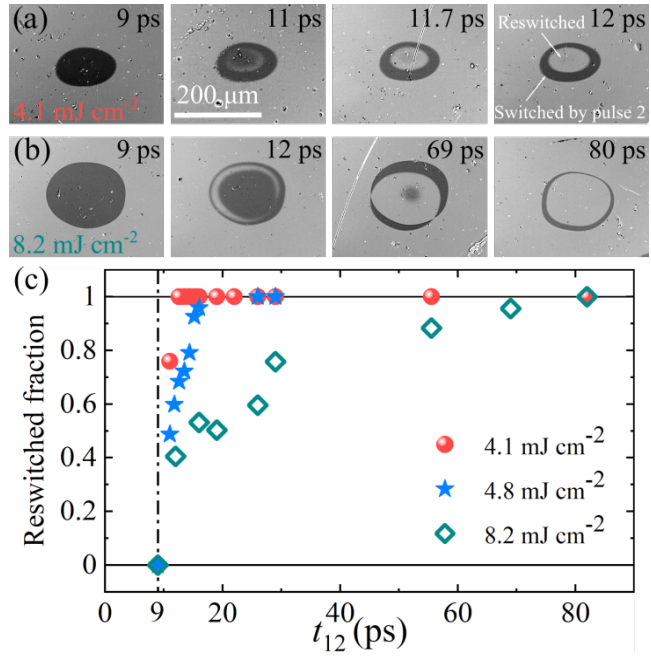

Figure 3: (a) Kerr micrographs of the irradiated region taken after dual pump excitation for different times $t_{12}$ separating the pulses. Both pump fluences are 4.1 mJ cm$^{-2}$. For $t_{12}$ = 12 ps, the bright center has been switched by the first pulse and toggled by the second, while the surrounding dark ring has been switched by the second pulse as described in the text. (b) Same as (a) but the first and second pump fluences are 8.2 mJ cm$^{-2}$ and 6 mJ cm$^{-2}$. (c) Variation of the re-switched fraction with $t_{12}$ for different pump fluences.

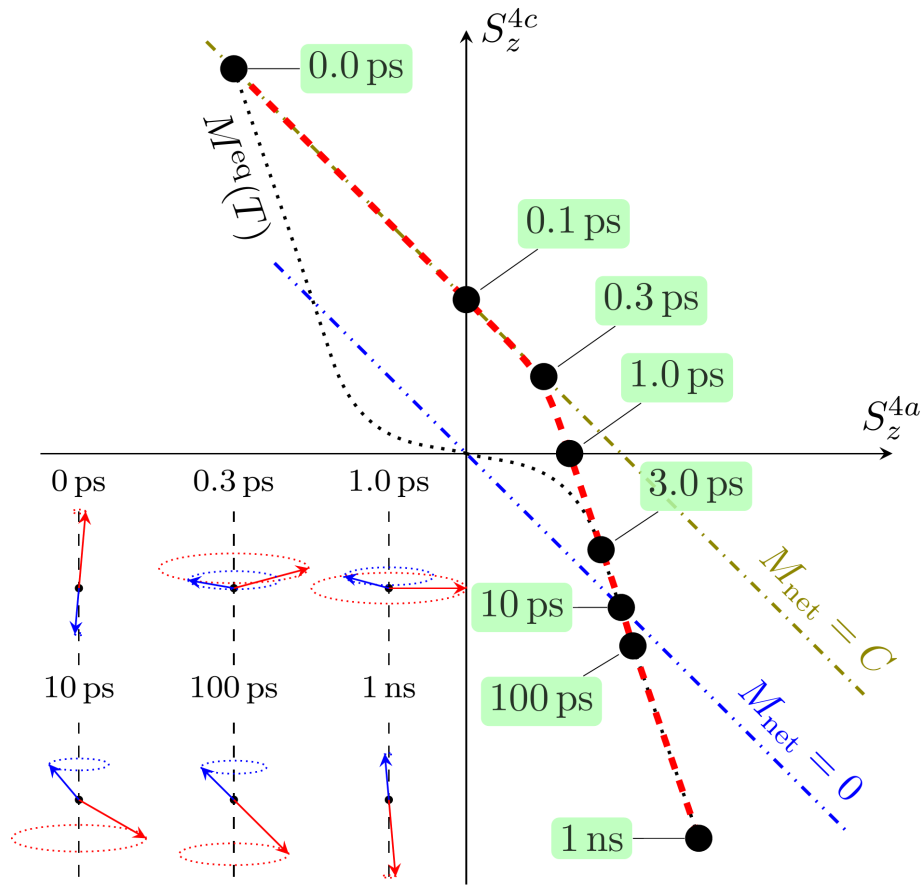

Figure 4: Magnetization of the 4*a* and 4*c* sublattices during SP-AOS trajectory (red dashed line). The first stage of exchange driven demagnetization switches the 4*a* sublattice while keeping the net magnetization constant; a transient ferromagnetic like state is reached between 0.1 ps and 0.3 ps. In the second stage, between 0.3 ps and 3.0 ps, the 4*c* sublattice reverses and the system relaxes toward equilibrium on the black dotted line. Between 3 ps and 1 ns, the system cools down. At 10 ps, the net magnetization changes sign and the system can be than be re-switched. The inset in the third quadrant illustrates the proposed spin configurations.